\documentclass[twocolumn]{aastex63}
\usepackage{amssymb,amsmath}
\usepackage{graphicx}
\usepackage{xcolor,natbib}
\usepackage{multirow}
\usepackage[T1]{fontenc}
\usepackage{ae,aecompl}
\usepackage{newtxtext, newtxmath}
\usepackage{float}
\usepackage{subfigure}
\usepackage{longtable}
\usepackage{enumitem}
\usepackage{longtable,listings}
\usepackage[flushleft]{threeparttable}
\usepackage{parcolumns}
\def\msig{$M_{\rm BH}- \sigma$\ }
\def\nii{[N~{\sc ii}]$\lambda6583$\AA}

\def\oiii{[O~{\sc iii}]$\lambda5007$\AA}
\def\obj{SDSS J1039}
\def\oii{[O~{\sc ii}]$\lambda3727$\AA\ }

\submitjournal{ApJ}
\accepted{Sep. 3rd, 2021}
\shorttitle{A candidate of true Type-2 AGN}
\shortauthors{Zhang et al.}

\begin{document}

\title{Long-term variability of the composite galaxy SDSS J103911-000057: A candidate of true Type-2 AGN}

\correspondingauthor{Zhang XueGuang}
\email{xgzhang@njnu.edu.cn}

\author{Zhang XueGuang}
\affiliation{School of physics and technology, Nanjing Normal University,
                No. 1, Wenyuan Road, Nanjing, Jiangsu, 210046, P. R. China}

\author{Zhang Yingfei}
\affiliation{School of physics and technology, Nanjing Normal University,
                No. 1, Wenyuan Road, Nanjing, Jiangsu, 210046, P. R. China}

\author{Chen PeiZhen}
\affiliation{School of physics and technology, Nanjing Normal University,
                No. 1, Wenyuan Road, Nanjing, Jiangsu, 210046, P. R. China}

\author{Wang BaoHan}
\affiliation{School of physics and technology, Nanjing Normal University,
                No. 1, Wenyuan Road, Nanjing, Jiangsu, 210046, P. R. China}

\author{Lv Yi-Li}
\affiliation{School of physics and technology, Nanjing Normal University,
                No. 1, Wenyuan Road, Nanjing, Jiangsu, 210046, P. R. China}

\author{Yu HaiChao}
\affiliation{School of physics and technology, Nanjing Normal University,
                No. 1, Wenyuan Road, Nanjing, Jiangsu, 210046, P. R. China}

\begin{abstract} 
	In the manuscript, the composite galaxy SDSS J103911-000057 (=\obj) is reported 
as a better candidate of true Type-2 AGN without hidden BLRs. None broad but only narrow 
emission lines detected in \obj~ can be well confirmed both by the F-test technique and 
by the expected broad emission lines with EW smaller than 13.5\AA~ with 99\% confidence 
level. Meanwhile, a reliable AGN power law component is preferred with confidence level 
higher than 7sigma in \obj. Furthermore, the long-term variability of \obj~ from 
CSS can be well described by the DRW process with intrinsic variability timescale 
$\tau\sim100{\rm days}$, similar as normal quasars. And, based on BH mass in \obj~ through 
the \msig relation and on the correlation between AGN continuum luminosity and total 
H$\alpha$ luminosity, the expected broad H$\alpha$, if there was, could be re-constructed 
with line width about $300-1000{\rm km~s^{-1}}$ and with line flux about 
$666\times10^{-17}{\rm erg~s^{-1}~cm^{-2}}$ under the Virialization assumption to BLRs, 
providing robust evidence to reject the probability that the intrinsic probable broad 
H$\alpha$ were overwhelmed by noises of the SDSS spectrum in \obj. Moreover, the \obj~ 
does follow the same correlation between continuum luminosity and [O~{\sc iii}] line 
luminosity as the one for normal broad line AGN, indicating \obj~ classified as a 
changing-look AGN at dim state can be well ruled out. Therefore, under the current knowledge, 
\obj~ is a better candidate of true Type-2 AGN. 
\end{abstract}


\section{Introduction}   

       Long-term variability tightly related to central BH (black hole) accreting process 
is one of fundamental characteristics of Type-1 AGN (broad line Active Galactic Nuclei) 
\citep{mr84, um97, ms16, bg20}. The long-term variability, especially in optical band, 
can be well modeled by the Continuous AutoRegressive (CAR) process firstly proposed by 
\citet{kbs09} and then the improved damped random walk (DRW) process in \citet{koz10, zk13, 
kb14, sh16, zk16}. Based on the CAR/DRW process described long-term variability properties, 
AGN selections have been well applied, such as the discussed results in \citet{mb11, kp11, 
cg14, tm17, sl19, db20}. As well as the long-term optical variability properties, broad 
optical emission lines coming from central broad emission line regions (BLRs) are another 
fundamental characteristics of Type-1 AGN \citep{sm00, ch11, bl18, zc19}. Combining the 
long-term variability with broad line emission features, Type-1 AGN can be well identified.

    Besides Type-1 AGN with apparent broad emission lines and also apparent long-term 
variability, there is one another kind of AGN: the Type-2 AGN with only narrow emission 
lines coming from narrow line regions (NLRs) in optical spectra. Properties of narrow 
emission lines have been well applied to detect central AGN activities in narrow 
emission-line galaxies, such as the results on the ongoing improved BPT diagrams 
\citep{bpt, kh03, gk06, kg06, csm11, jbc14, ks17, kn19, zh20}. Moreover, based on the 
discovery of polarized broad permitted lines in \citet{am85, tr01, tr03, nk04, sg18} in 
some Type-2 AGN, the well-known Unified Model in \citet{an93,nh15, aa17} has been proposed 
and well accepted to explain the different observational optical phenomena mainly due 
to orientation effects of central dust torus. Therefore, under the framework of the 
Unified model, Type-1 AGN and Type-2 AGN have the same central geometric structures, 
but central engine and BLRs hidden in Type-2 AGN. 

    However, based on the interesting work on detecting polarized broad emission lines in 
Type-2 AGN, especially on the pioneer work in \citet{tr01}, there is one special kind of 
AGN: true Type-2 AGN (or Type-2 AGN with none hidden-BLRs, or BLRs-less AGN, or unobscured 
Type-2 AGN, or naked AGN in literature), with no expected hidden central BLRs. \citet{tr01, 
tr03} have shown spectropolarimetric results of about 30 Seyfert 2 galaxies with no hidden 
BLRs, due to lack of polarized broad emission lines. \citet{sg10} have reported two strong 
candidates of true Type-2 AGN, NGC3147 and NGC4594, which have few X-ray extinctions and 
have the upper limits on the broad emission line luminosities are two orders of magnitude 
lower than the average of typical Type-1 AGN, leading to the conclusion that "true Type-2 
AGN do exist but they are very rare". \citet{bv14} have shown that there is possibility 
that a very small number of Type-2 quasar with strong g-band variability might be true 
Type-2 AGN. \citet{zh14} have shown a sample of candidates of true Type-2 AGN with both 
the long-term variability and the expected reliable power law components in their spectra 
in SDSS (Sloan Digital Sky Survey). \citet{ly15} have reported a candidate of true Type-2 
AGN, SDSS J0120, with long-term variability but none-detected broad emission lines. 
More recently, \citet{pw16} have shown a sample of candidates of true Type-2 AGN based 
on properties of unobscured X-ray emissions.

     Studying on true Type-2 AGN can provide further clues on formation or suppression 
of central BLRs in AGN, such as the proposed theoretical models on the nature of true 
Type-2 AGN in \citet{nm03, eh09, cao10, en16}, strongly indicating disappearance of 
central BLRs depends on physical properties of central AGN activities and/or depends on 
properties of central dust obscuration. \citet{nm03} have shown that "the absence or 
presence of central BLRs can be well regulated by accretion rate, assumed that the BLRs 
are formed by accretion disk instabilities occurring in proximity of the critical radius 
at which the disk changes from gas pressure dominated to radiation pressure dominated". 
\citet{eh09} have proposed a disk-wind scenario in AGN to predict the disappearance of 
the central BLRs at luminosities lower than $5\times10^{39}M_7^{2/3}{\rm erg~s^{-1}}$ with 
$M_7$ as central BH mass in unit of $10^7{\rm M_\odot}$. \citet{cao10} has shown that 
"disappearance of central BLRs associated with the outflows from the accretion disks can 
be expected in AGN with Eddington ratio smaller than 0.001, because the inner small cold 
disk is evaporated completely in the advection dominated accretion flows (ADAF) and outer 
thin accretion disk may be suppressed by the ADAF". \citet{en16} have shown that "true 
Type-2 AGN should have higher luminosities than $4\times10^{46}{\rm erg~s^{-1}}$ 
considering mass conservations in the disk outflows".

      Commonly, disappearance of broad emission lines could also be probably due to lower 
spectral quality, such as the results in Mrk573 and NGC3147. Mrk573 has been classified 
as a true Type-2 AGN in \citet{tr01}. However, \citet{nk04} have shown the clearly detected 
polarized broad Balmer lines, leading Mrk573 as a normal Type-2 AGN with central hidden 
BLRs. Similarly, NGC3147 has been previously classified as a true Type-2 AGN in \citet{sg10, 
ba12} through both spectropolarimetric results and unobscured X-ray emission properties, 
however \citet{ba19} more recently have clearly detected double-peaked broad H$\alpha$ in 
high quality HST spectrum. Meanwhile, \citet{ip15} have shown that none detected polarized 
broad emission lines in candidates of some true Type-2 AGN are probably due to effects 
of less scattering medium due to the reduced scattering volume given the small torus 
opening angle and/or due to effects of the increased torus obscurations.

     There is so-far no definite conclusions on the very existence of true Type-2 AGN, 
neither clear conclusion on the physical nature of true Type-2 AGN, but more candidates 
of true Type-2 AGN can provide further clues on intrinsic physical properties of true 
Type-2 AGN. In the manuscript, we report one new candidate of true Type-2 AGN in SDSS 
J103911-000057 (=\obj) through both its apparent long-term optical variability and 
high-quality spectroscopic features having no broad emission lines. The manuscript is 
organized as follows. Section 2 presents our main results on spectroscopic emission line 
features and the properties of optical long-term V-band variability of \obj. Necessary 
discussions are shown in Section 3. Section 4 gives the final conclusions. And in this 
manuscript, we have adopted the cosmological parameters of 
$H_{0}=70{\rm km\cdot s}^{-1}{\rm Mpc}^{-1}$, $\Omega_{\Lambda}=0.7$ and $\Omega_{\rm m}=0.3$.

\begin{figure*}[htp]   
\centering\includegraphics[width = 18cm,height=6cm]{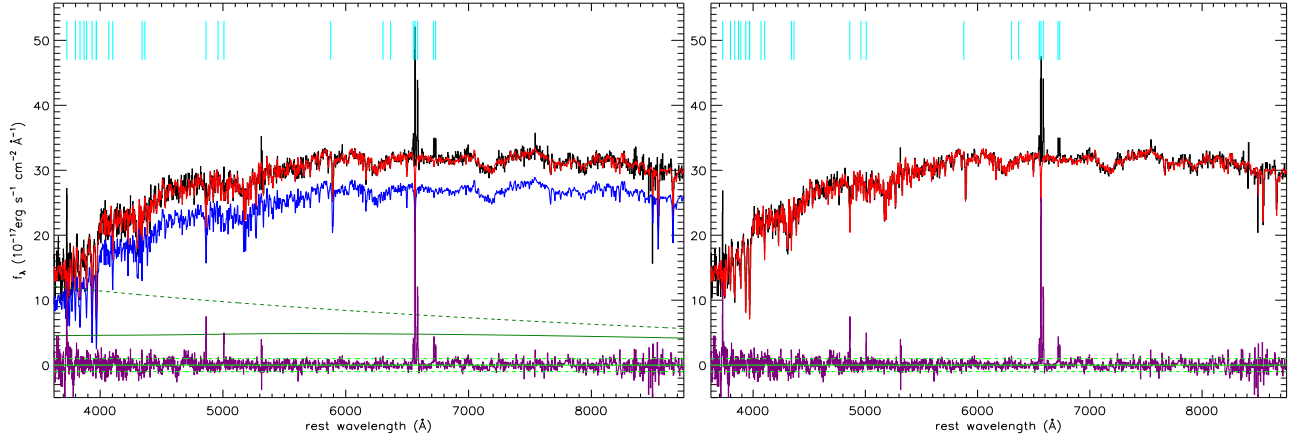}
\caption{SDSS spectrum of \obj~ and the determined host galaxy contributions by the 
SSP method with (in left panel) and without (in right panel) considerations of a power 
law component. In each panel, from top to bottom, solid black line shows the observed SDSS 
spectrum, solid red line shows the determined best descriptions to the spectrum with emission 
lines being masked out, solid purple line shows the line spectrum after subtractions of the 
best descriptions, horizontal solid green line shows $f_\lambda=0$, and horizontal dashed 
green lines show $f_\lambda=\pm1$, respectively. In the left panel, solid blue line shows 
the determined stellar lights of host galaxy, solid dark green line shows the determined 
reddened AGN power law component, dashed dark green line shows the reddening corrected AGN 
power component. In each panel, from left to right, the vertical cyan lines mark the following 
emission features which are being masked out when the SSP method is running, including \oii, 
H$\theta$, H$\eta$, [Ne~{\sc iii}]$\lambda3869$\AA, He~{\sc i}$\lambda3891$\AA, Calcium K line, 
[Ne~{\sc iii}]$\lambda3968$\AA, Calcium H line, [S~{\sc ii}]$\lambda4070$\AA, H$\delta$, 
H$\gamma$, [O~{\sc iii}]$\lambda4364$\AA, H$\beta$, [O~{\sc iii}]$\lambda4959,5007$\AA, 
He~{\sc i}$\lambda5877$\AA, [O~{\sc i}]$\lambda6300,6363$\AA, [N~{\sc ii}]$\lambda6548$\AA, 
H$\alpha$, [N~{\sc ii}]$\lambda6583$\AA, and [S~{\sc ii}]$\lambda6716,6732$\AA, respectively.}
\label{spec}
\end{figure*}

\begin{figure*}[htp]  
\centering\includegraphics[width = 18cm,height=6cm]{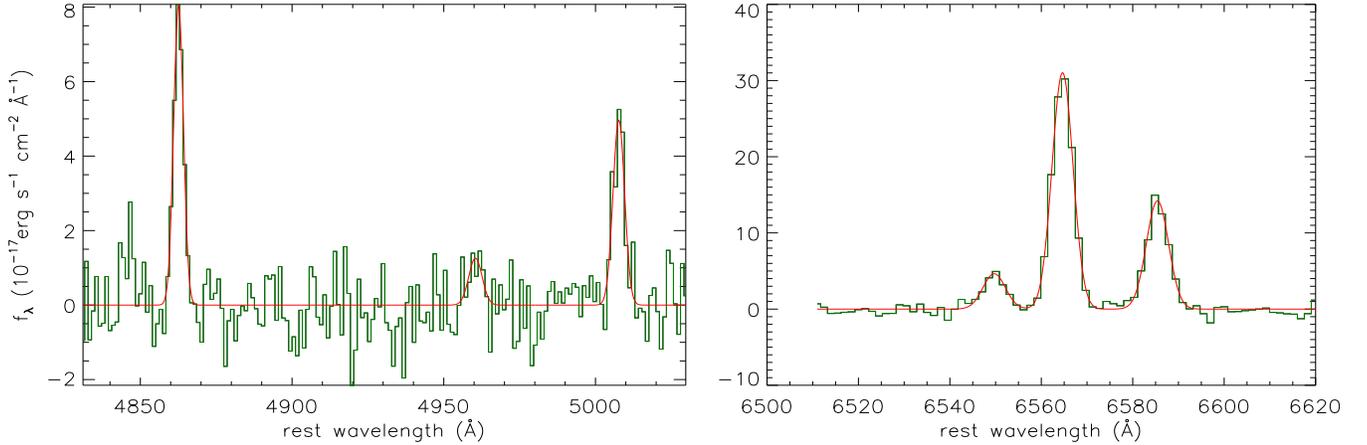}
\caption{The Gaussian fitted results to the emission lines around H$\beta$ (left 
panel) and around H$\alpha$ (right panel). In each panel, solid line in dark green 
shows the line spectrum, and solid red line shows the best fitting results.}
\label{line}
\end{figure*}

\begin{figure}[htp]  
\centering\includegraphics[width = 8cm,height=7cm]{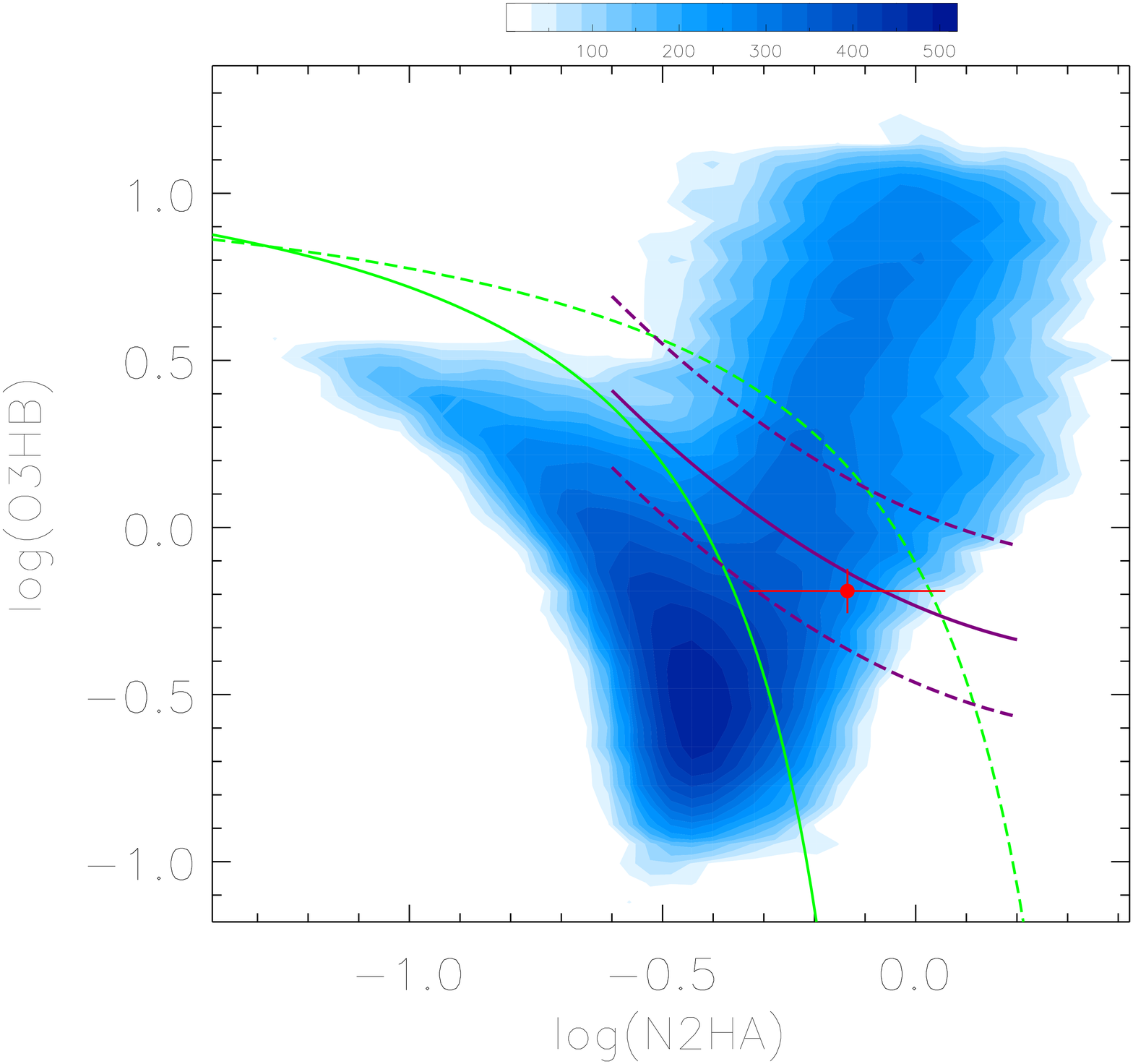}
\caption{Properties of \obj~ (solid red circle plus error bars) in the BPT diagram 
of O3HB versus N2HA. Solid and dashed green lines show the dividing lines reported in 
\citet{kh03} and in \citet{ke01} between HII galaxies, composite galaxies and AGN. Solid 
purple line and dashed purple lines show the dividing line and area for composite galaxies 
determined in our recent work in \citet{zh20}. The contour is created by emission line 
properties of more than 35000 narrow emission-line galaxies discussed in \citet{zh20} 
collected from SDSS DR15. Corresponding number densities to different colors are shown 
in the color bar.}
\label{bpt}
\end{figure}

\section{Spectroscopic and long-term photometric variability properties of \obj}

   \obj~ at redshift 0.0499 is a main galaxy in SDSS DR16 \citep{aa20}, with the apparent 
optical Petroson magnitudes of 20.05, 18.01, 17.21, 16.78, 16.46 at u, g, r, i and z bands, 
respectively. And based on the SDSS provided inverse concentration indices (ratio of the 
two Petrosian radii $r50/r90$ with $r50$ as half light radius and $r90$ as 90\% light radius) 
of 0.67, 0.61, 0.60, 0.59, 0.60 at u, g, r, i and z bands, \obj~ is a late-type galaxy, 
based on the discussed results in \citet{si01, sf01} (see also descriptions in 
\url{https://www.sdss.org/dr12/algorithms/classify/}). Fig.~\ref{spec} shows the galactic 
reddening corrected spectrum (PLATE-MJD-FIBERID=0274-51913-0232) of \obj~ observed in January 
2001 with total exposure time of 5400 seconds and with median signal-to-noise $SN\sim34$. 

     It is clear that the SDSS spectrum of \obj~ has apparent host galaxy contributions. 
In order to check whether is there an AGN component and in order to well measure emission 
line properties, host galaxy contributions should be firstly determined. The commonly accepted 
SSP (Simple Stellar Population) method is well applied. The SSP method has been firstly proposed 
in \citet{bc03}, to describe observed spectra of galaxies by stellar population synthesis. 
Then, the SSP method has been widely applied to determined the star formation history, 
metallicity and dust content of galaxy, such as the detailed discussions in \citet{kh03, 
cm05, cm17}, etc. Here, the 39 simple stellar population templates from \citet{bc03} have 
been exploited, which can be used to well-describe the characteristics of almost all the 
SDSS galaxies as discussed in \citet{bc03}. Meanwhile, there is an additional component, a 
power law component $\alpha\times\lambda^\beta$, applied to describe the intrinsic AGN 
continuum emissions which can be well confirmed by the following shown long-term variability. 
Moreover, the intrinsic reddening effects can be well considered by a parameter of $E(B-V)$. 
Then, similar as what we have done in \citet{zh14, zh19, zh21a, zh21b}, with the emission 
lines listed in \url{http://classic.sdss.org/dr1/algorithms/speclinefits.html#linelist} 
being masked out by full width at zero intensity about 450${\rm km~s^{-1}}$, the observed 
SDSS spectrum of \obj~ can be well described by the broadened and reddened SSPs (the 
broadening velocity as the stellar velocity dispersion) plus the reddened power law component 
through the Levenberg-Marquardt least-squares minimization technique. 

	When the model functions are applied, there are 43 model parameters, 39 strengthen 
factors with zero as the starting values for the 39 SSPs, the broadening velocity $\sigma_\star$ 
with ${\rm 100km~s^{-1}}$ as the starting value, $\alpha$ and $\beta$ for the power law 
component with zero as the starting values, and the parameter of $E(B-V)$ with zero as the 
starting value. Then, the best descriptions to the spectrum of \obj~ can be well determined, 
leading the determined $\chi^2/Dof$ (the summed squared residuals divided by degree of freedom) 
to be about $\chi^2_2/Dof_2\sim3925.5/3561$, and the determined stellar velocity dispersion 
to be about $\sigma_\star\sim76.5\pm4.3{\rm km~s^{-1}}$, and the determined parameter 
$E(B-V)$ to be about 0.22, and the determined reddened power law component described by 
$4.67\times(\lambda/5100\text{\AA})^{-0.04}$. The best-fitting results are shown in the 
left panel of Fig.~\ref{spec}. Based on the well determined host galaxy contributions 
and the power law component, the continuum intensity ratio at 5100\AA~ is about 4.6 of 
the continuum emissions from stellar lights to the AGN emissions.

    Moreover, the commonly accepted F-test technique is applied to determine whether 
the fitting procedure above determined power law component is necessary enough. Without 
considering the power law component, only the broadened SSPs are applied to describe the 
observed SDSS spectrum of \obj~ with emission lines being masked out. Through the 
Levenberg-Marquardt least-squares minimization technique, the best descriptions to the 
SDSS spectrum of \obj~ by only the SSPs are shown in the right panel of Fig.~\ref{spec}, 
leading the determined $\chi^2/Dof$ to be about $\chi^2_1/Dof_1\sim3953.6/3563$, and 
the determined stellar velocity dispersion to be about 
$\sigma_\star\sim81.2\pm4.1{\rm km~s^{-1}}$, and the determined $E(B-V)$ to be about 
0.26. Based on the different $\chi^2/Dof$ for the different model functions with and 
without considerations of the power law component, the calculated $F_p$ value is about 
\begin{equation}
F_p=\frac{\frac{\chi^2_1-\chi^2_2}{Dof_1-Dof_2}}{\chi^2_2/Dof_2}\sim12.75
\end{equation}. 
Based on $Dof_1-Dof_2$ and $Dof_2$ as number of Dofs of the F distribution numerator and 
denominator, the expected value from the statistical F-test with confidence level higher 
than 99.9997\% will be near to $F_p$. Therefore, the power law component can be well 
accepted with confidence level higher than 99.9997\% (higher than 7sigma).

\begin{table}
\caption{Line parameters of emission lines of \obj}
\begin{tabular}{lccc}
\hline\hline
Line    &   $\lambda_0$  &   $\sigma$  & flux  \\
\hline
H$\alpha$ & 6564.65$\pm$0.06 & 2.29$\pm$0.05 & 178.6$\pm$3.6 \\
H$\beta$  & 4862.59$\pm$0.14 & 1.61$\pm$0.14 & 32.6$\pm$2.5 \\
\oiii & 5007.68$\pm$0.27 & 1.92$\pm$0.26 & 23.9$\pm$2.9 \\
\nii & 6585.41$\pm$0.11 & 2.43$\pm$0.11 & 86.7$\pm$3.3 \\
\hline
\end{tabular}\\
Notice: The second, third and fourth columns show the central wavelength in unit of \AA\ 
in rest frame, the line width (second moment) in unit of \AA\ and the line flux in unit 
of $10^{-17}{\rm erg~s^{-1}~cm^{-2}}$.
\end{table}

     After subtractions of the stellar lights and the power law component shown in the 
left panel of Fig.~\ref{spec}, the emission lines in the line spectrum can be well measured 
by narrow Gaussian functions (second moment smaller than 400${\rm km~s^{-1}}$). There are 
three narrow Gaussian functions are applied to describe the narrow H$\beta$ and 
[O~{\sc iii}]$\lambda4959,5007$\AA~ doublet within the rest wavelength from 4830\AA~ to 
5030\AA, and three narrow Gaussian functions are applied to describe the narrow H$\alpha$ 
and [N~{\sc ii}]$\lambda6548,6583$\AA~ doublet within the rest wavelength from 6510\AA~ 
to 6620\AA. Then, through the Levenberg-Marquardt least-squares minimization technique, 
the best fitting results to the emission lines are shown in Fig.~\ref{line} with the 
determined $\chi^2/Dof$ to be about $\chi^2_1/Dof_1\sim200.446/231$. When the model 
functions are applied, three parameters of each Gaussian function are free parameters, 
and there are no further restrictions on the model parameters besides the restrictions 
that line fluxes are not smaller than zero.

	Why it is not necessary to consider broad Balmer components? We answer the question 
by the F-test statistical technique. Besides the pure narrow Gaussian functions applied above, 
two broad Gaussian functions (second moment larger than 400${\rm km~s^{-1}}$) with central 
wavelengths fixed to 4862\AA~ and 6564\AA~ are applied to describe the probable broad H$\beta$ 
and broad H$\alpha$. Then, within the rest wavelength from 4830\AA~ to 5030\AA~ and from 
6510\AA~ to 6620\AA, the emission lines in the line spectrum are described again by the new 
model functions through the Levenberg-Marquardt least-squares minimization technique, leading 
the determined $\chi^2/Dof$ to be about $\chi^2_2/Dof_2\sim199.758/227$. Then, the calculated 
$F_p$ value is about
\begin{equation}
F_p=\frac{\frac{\chi^2_1-\chi^2_2}{Dof_1-Dof_2}}{\chi^2_2/Dof_2}\sim0.195
\end{equation}.
Based on $Dof_1-Dof_2=4$ and $Dof_2=227$ as number of Dofs of the F distribution numerator 
and denominator, the expected value from the statistical F-test with confidence level around 
6\% will be near to $F_p$. Therefore, we can safely accepted that it is not necessary to 
consider the broad Balmer emission components in \obj, due to the lower confidence level 
around 6\%.

	The measured line parameters are listed in Table~1. The flux ratios of \oiii~ 
to narrow H$\beta$ (O3HB) and of \nii~ to narrow H$\alpha$ (N2HA) are about 0.73 and 
0.64, respectively, indicating \obj~ is a standard composite galaxy, based on the dividing 
lines reported in \citet{ke01, kh03, kg06} and reported regions for composite galaxies 
in \citet{zh20}. Fig.~\ref{bpt} shows properties of \obj~ in the BPT diagram of N2HA versus 
O3HB, leading \obj~ to be a well classified composite galaxy with a weak AGN component 
with confidence levels higher than 7sigma. In order to confirm \obj~ as a true Type-2 AGN, 
variability properties can provide further and robust evidence, combining with the 
spectroscopic properties.


   There are four reasonable explanations on the spectroscopic features having no broad emission 
lines of one narrow emission-line object. First, the narrow emission-line object is a normal Type-2 
AGN with hidden central BLRs. Second, the narrow emission-line object is a true Type-2 AGN with lack 
of central BLRs. Third, the narrow emission-line object is a quiescent galaxy with no considerations 
of broad emission lines. Fourth, the narrow emission-line object is an AGN but with central BLRs 
seriously obscured. Long-term variability, as one of fundamental characteristics of broad line AGN 
(no variability in quiescent galaxies), will provide robust evidence to confirm which explanation is 
preferred. If there was apparent long-term variability indicating the central engine of AGN can be 
directly observed, both the second and the fourth explanation could be preferred. Certainly, if 
there was no apparent long-term variability, it would be hard to confirm which explanation is preferred. 
In the \obj, the seriously obscured central BLRs can also be well ruled out, because the confirmed 
apparent AGN power law component should lead to expected apparaent broad emission lines as well 
discussed re-constructed broad Balmer lines in the subsection 3.1, against the spectroscopic features 
with no broad emission lines. Therefore, if there was apparent long-term variability in \obj, only 
the second explanation can be well preferred, leading to the conclusion of lack of central BLRs in \obj.

\begin{figure}[htp]
\centering\includegraphics[width = 8cm,height=10cm]{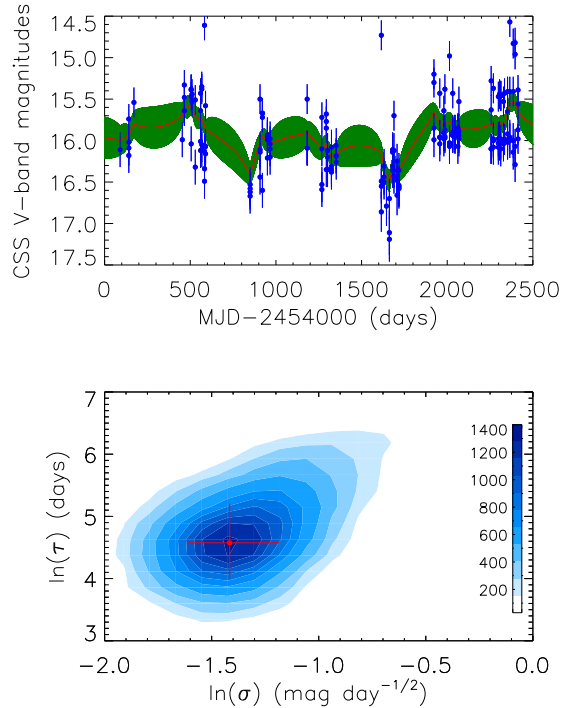}
\caption{Top panel shows the CSS V-band light curve shown as solid blue circles plus error 
bars of \obj~ and the determined best descriptions shown as solid red line by the public 
JAVELIN code. In top panel, area covered by dark green shows the corresponding 1sigma 
confidence bands to the best descriptions. Bottom panel shows the MCMC determined 
two-dimensional projected posterior distributions of the two DRW parameters of $\sigma$ 
in unit of ${\rm mag~day^{-0.5}}$ and $\tau$ in unit of days. In bottom panel, solid circle 
plus error bars show the accepted values and corresponding uncertainties.}
\label{lmc}
\end{figure}

	The 6.4years-long photometric V-band light curve of \obj~ can be collected from CSS 
(Catalina Sky Survey) \citep{dr09} and shown in Fig.~\ref{lmc}. Then, the widely accepted DRW/CAR 
process is applied to check the variability properties of \obj, because the DRW/CAR process 
has been proved to be a preferred modeling process to describe AGN intrinsic variability, 
such as the well discussed results in \citet{mi10, bj12, ak13, zk13}. \citet{kbs09} have 
firstly proposed the CAR process \citep{bd02} to describe the AGN intrinsic variability, and 
found that the AGN intrinsic variability timescales are consistent with disk orbital or thermal 
timescales. \citet{koz10} have provided an improved robust mathematic method to estimate the DRW 
process parameters, and found that AGN variability could be well modeled by the DRW process. 
Then, \citet{zk11} have provided a public code of JAVELIN 
(\url{http://www.astronomy.ohio-state.edu/~yingzu/codes.html}) (Just Another Vehicle for Estimating 
Lags In Nuclei) based on the method in \citet{koz10}. Here, the JAVELIN code is accepted to 
describe the long-term variability of \obj. Through the MCMC (Markov Chain Monte Carlo, \citet{fh13}) 
analysis with the uniform logarithmic priors of the DRW process parameters of $\tau$ in unit of days 
and $\sigma$ in unit of ${\rm mag~day^{-0.5}}$ covering every possible corner of the parameter space 
($0~<~\tau/{\rm days}~<~1e+5$ and $0~<~\sigma/({\rm mag~day^{-0.5}})~<~1e+2$), the posterior 
distributions of the DRW process parameters can be well determined, similar as what we have done 
in \citet{zh17a}. The best descriptions to the light curve of \obj~ by the JAVELIN code are shown 
in the top panel of Fig.~\ref{lmc}. And the MCMC determined two-dimensional projected posterior 
distributions of $\ln(\sigma/({\rm mag~day^{-0.5}}))$ and $\ln(\tau/{\rm days})$ are shown in the 
bottom panel of Fig.~\ref{lmc}, with accepted $\ln(\sigma/({\rm mag~day^{-0.5}}))\sim-1.41_{-0.12}^{+0.31}$ 
and $\ln(\tau/{\rm days})\sim4.57_{-0.39}^{+0.81}$ ($\tau\sim97{\rm days}$). Comparing with the 
reported values of $\tau/{\rm days}$ in SDSS quasars in \citet{mi10} and in the sample of quasars in 
\citet{kbs09}, $\tau\sim97{\rm days}$ in \obj~ is a common value in quasars, indicating the 
long-term variability is connected to intrinsic AGN activities in \obj.

	Based on the DRW process described long-term variability, an intrinsic AGN component 
can be preferred, which also can be supported by the detected AGN power law components in the 
SDSS spectrum. Therefore, combining variability properties with the spectroscopic properties 
without broad emission lines, \obj~ can be well accepted as a candidate of true Type-2 AGN.

\section{Main discussions}

\begin{figure*}[htp]   
\centering\includegraphics[width = 15cm,height=10cm]{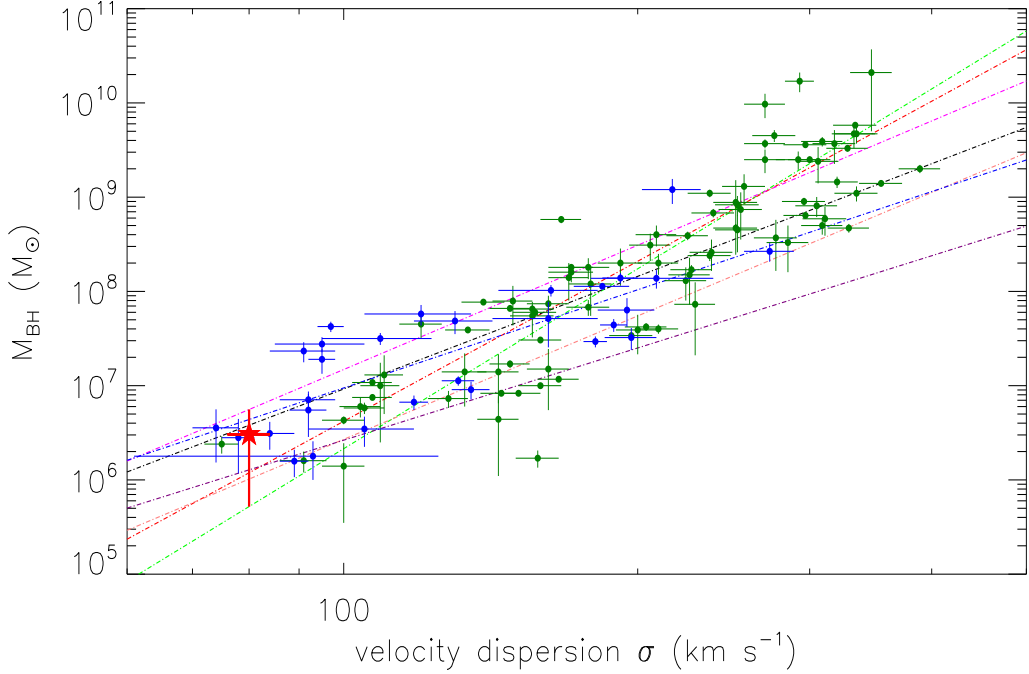}
\caption{On the correlation between stellar velocity dispersion and BH mass for both quiescent 
galaxies and AGN. Solid circles in dark green and in blue show the measured results for the 
89 quiescent galaxies well discussed in \citet{sg15} and the 29 RM (reverberation-mapped) AGN  
well discussed in \citet{wy15}, respectively. Dot-dashed lines in green, in red, in magenta, 
in black, in pink, in purple and in blue represent the reported \msig relations determined 
through the quiescent galaxies in \citet{sg15}, in \citet{mm13}, in \citet{kh13} and through 
the RM AGN in \citet{wy15}, the RM AGN with classical bulges in \citet{hk14}, the RM AGN  
with pseudobulges in \citet{hk14} and the RM AGN in \citet{ws13}, respectively. And the 
solid five-point-star in red shows the measured results of the \obj~ with BH mass about 
$(3.1\pm2.5)\times10^{6}{\rm M_\odot}$ and with the measured $\sigma_\star\sim80{\rm km~s^{-1}}$. 
} 
\label{msig}
\end{figure*}

\begin{figure}[htp]  
\centering\includegraphics[width = 9cm,height=5.5cm]{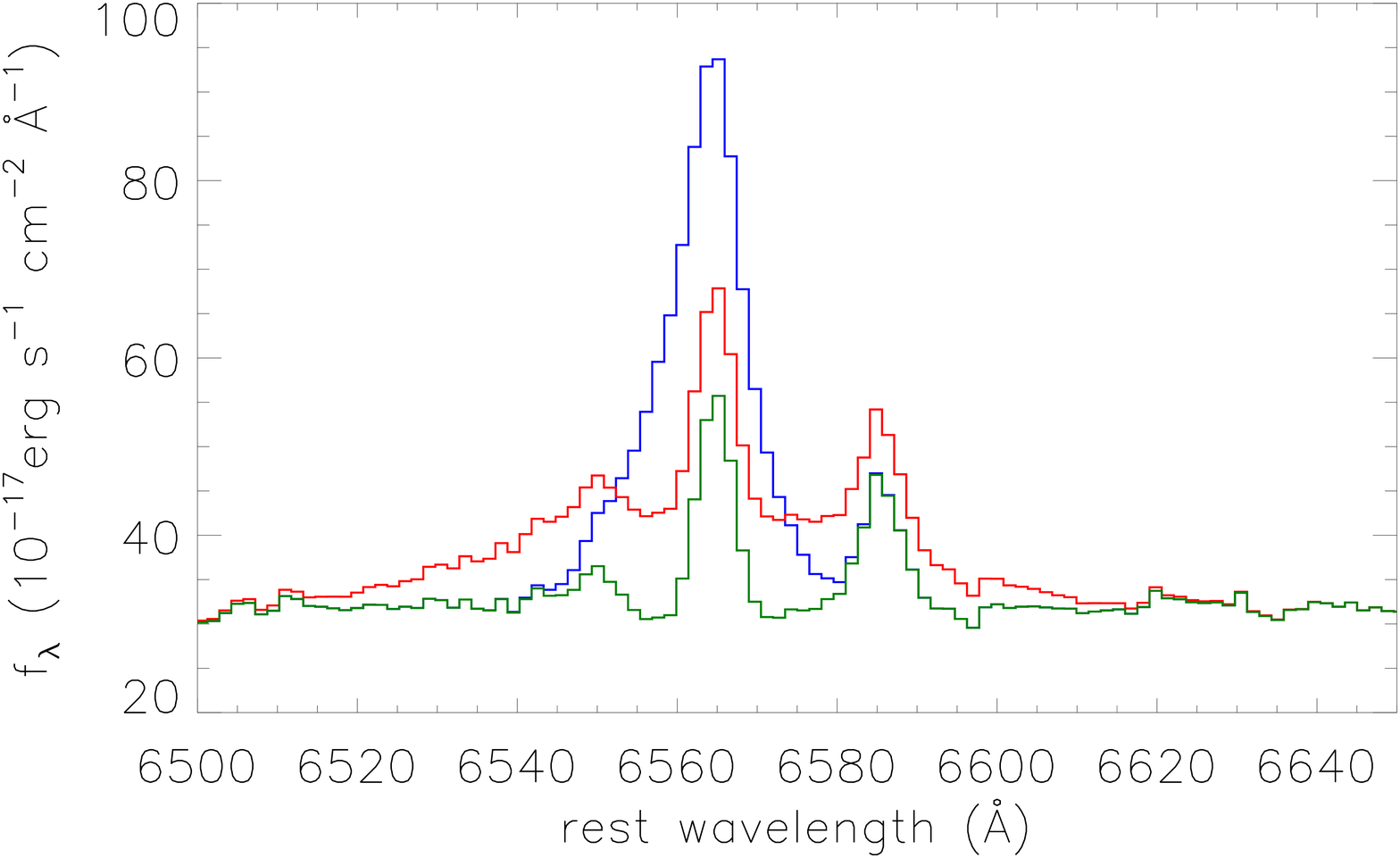}
\caption{The expected spectroscopic properties around H$\alpha$ with considering the re-constructed 
broad H$\alpha$ through the Virialization assumptions to BLRs. Solid line in dark green shows the 
SDSS spectrum of \obj~ within rest wavelength from 6500\AA~ to 6650\AA. Solid blue line 
shows the expected spectroscopic properties, if there was the expected broad H$\alpha$ re-constructed 
with second moment of $303{\rm km~s^{-1}}$ and line flux of $666\times10^{-17}{\rm erg~s^{-1}~cm^{-2}}$. 
Solid red line shows the expected spectroscopic properties, if there was the expected broad 
H$\alpha$ re-constructed with second moment of $996{\rm km~s^{-1}}$ and line flux of 
$666\times10^{-17}{\rm erg~s^{-1}~cm^{-2}}$.
}
\label{fake}
\end{figure}

\begin{figure}[htp]  
\centering\includegraphics[width = 9cm,height=6cm]{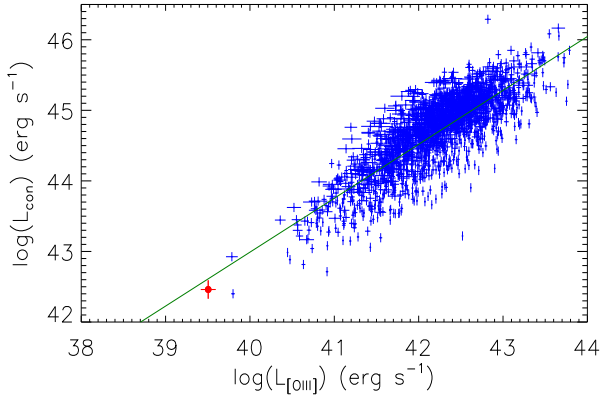}
\caption{On the correlation between continuum luminosity and total [O~{\sc iii}] line luminosity. 
Solid blue dots plus error bars are the results for Type-1 AGN discussed in \citet{zh17}. Solid 
red circle plus error bars are the results for the \obj.
}
\label{co3}
\end{figure}

\begin{figure}[htp] 
\centering\includegraphics[width = 9cm,height=6cm]{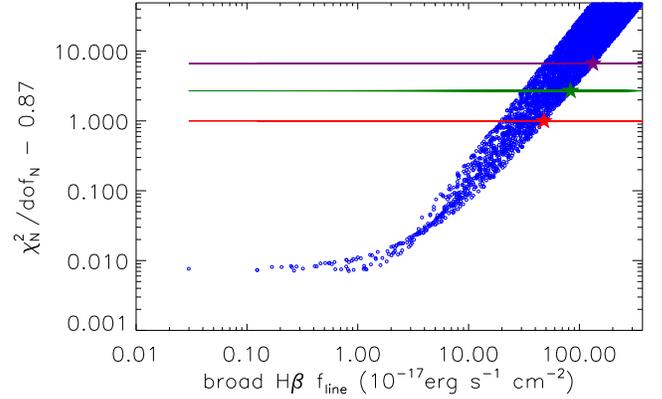}
\caption{Dependence of $\chi^2_N/dof_N - 0.87$ on the input line flux $f_{line}$ of expected 
broad H$\beta$. Open blue circles are the results from 5000 randomly collected values of 
$\sigma_{line}$ and $f_{line}$. The horizontal lines in red, in green and in purple show 
$\chi^2_N/dof_N - 0.87=1$ (68\% confidence level), $\chi^2_N/dof_N - 0.87=2.7$ (90\% confidence 
level) and $\chi^2_N/dof_N - 0.87=6.63$ (99\% confidence level), respectively. The five-point-stars 
in red, in green and in purple show the upper limits of $f_{line}$ with confidence levels of 
68\%, 90\% and 99\%, respectively.}
\label{chi2}
\end{figure}

\begin{table}
\caption{The reported \msig relations in literature}
	\begin{center}\begin{tabular}{lcccc}
\hline\hline
obj    &   N  &   $\alpha_{BH}$  & $\beta_{BH}$  & Ref \\
\hline
QG  &  89  & 8.24$\pm$0.10 & 6.34$\pm$0.80 & \citet{sg15} \\
QG  &  72  & 8.32$\pm$0.05  & 5.64$\pm$0.32 & \citet{mm13} \\
QG  & 51  &  8.49$\pm$0.05  & 4.38$\pm$0.29 & \citet{kh13} \\
RA  & 29 & 8.16$\pm$0.18 & 3.97$\pm$0.56 & \citet{wy15} \\
RAC & 16 & 7.74$\pm$0.13 & 4.35$\pm$0.58 & \citet{hk14} \\
RAP & 14 &  7.40$\pm$0.19 & 3.25$\pm$0.76 & \citet{hk14} \\
RA  & 25 & 8.02$\pm$0.15 & 3.46$\pm$0.61 & \citet{ws13} \\
\hline
\end{tabular}\\
\end{center}
Notice: The first column shows what objects are used to determine the \msig relation: 
QG for Quiescent galaxies, RA for reverberation mapped AGN, RAC for reverberation 
mapped AGN with classical bulges, RAP for reverberation mapped AGN with pseudobulges. 
The second column shows number of the used objects. The third and the forth columns 
list the values of $\alpha_{BH}$ and $\beta_{BH}$, respectively. The fifth column 
shows the corresponding reference.
\end{table}

\subsection{Properties of re-constructed broad emission lines, if there were broad emission lines 
overwhelmed in the SDSS spectrum}

	In the subsection, two methods are applied to determine whether are there expected 
broad emission lines overwhelmed in the SDSS spectrum of the \obj. One method is to re-construct 
broad H$\alpha$ after considering Virialization assumptions to BLRs. The other method is applied 
to determine the upper limits of line flux of the expected broad emission lines in the \obj.

	The first method is applied as follows. Based on the measured stellar 
velocity dispersion about $\sigma_\star\sim80{\rm km/s}$, the BH mass can be estimated 
through the well-known \msig relation, 
\begin{equation}
\log(\frac{M_{\rm BH}}{\rm M_\odot})~=~\alpha_{BH}~+~\beta_{BH}~\times~
	\log(\frac{\sigma}{\rm 200~km~s^{-1}})
\end{equation}. 
The \msig relation is firstly reported in \citet*{fm00, ge00}, based on dynamic measured 
BH masses and measured stellar velocity dispersions of a small sample of nearby quiescent 
galaxies. More recent review of the \msig relation for galaxies can be found in \citet{kh13}. 
And now there are plenty of studies on the \msig relation for both quiescent galaxies and AGN, 
such as the results well discussed in \citet{mm13, ws13, hk14, sg15, wy15, bb17}, leading to 
reported \msig relations with different slopes which are listed in Table~2 and shown in 
Fig.~\ref{msig}. Based on the measured stellar velocity dispersion in \obj, minimum and maximum 
BH mass of \obj~ can be simply estimated through the different \msig relations. The \msig relation 
reported in \citet{kh13} with $\alpha_{BH}=8.49$ and $\beta_{BH}=4.38$ (shown as dot-dashed line 
in magenta in Fig.~\ref{msig}) is applied to determine $5.6\times10^6{\rm M_\odot}$ as 
the maximum BH mass of \obj. And the \msig relation in \citet{sg15} with $\alpha_{BH}=8.24$ 
and $\beta_{BH}=6.34$ (shown as dot-dashed line in green in Fig.~\ref{msig}) is applied to 
determine $5.2\times10^5{\rm M_\odot}$ as the minimum BH mass of \obj. The results are shown 
in Fig.~\ref{msig} for \obj~ with BH mass $(3.1\pm2.5)\times10^{6}{\rm M_\odot}$. Then based 
on the determined reddening corrected AGN continuum emissions shown as dashed line in dark 
green in the left panel of Fig.~\ref{spec}, the intrinsic continuum luminosity at 5100\AA\ is 
about $L_{con}\sim{\rm 2.89\times10^{42}erg~s^{-1}}$, leading to the expected BLRs size about 
$R_{BLRs}\sim5.26{\rm light-days}$ through the more recent empirical relation 
$R_{BLRs}\propto L_{con}^{0.554}$ in \citet{bd13}.

	Then, under the virialization assumption to broad line emission clouds in BLRs 
\citep{pf04, sh11, rh11}, virial BH mass can be estimated by 
\begin{equation}
M_{BH}=5.5\times\frac{R_{BLRs}\times\sigma_{broad}^2}{G}
\end{equation},
where $\sigma_{broad}$ means the second moment (line width) of the broad Balmer emission 
lines. The lower and upper limits of line width $\sigma_{broad}$ of broad H$\alpha$ can be 
well estimated from ${\rm 303km~s^{-1}}$ to ${\rm 996km~s^{-1}}$ through the estimated minimum 
and maximum BH masses [$5.2\times10^5$, $5.6\times10^6$]${\rm M_\odot}$ of \obj~ through the 
\msig relation. Meanwhile, based on the strong linear correlation between continuum luminosity 
at 5100\AA~ ($L_{con}$) and line luminosity of both broad and narrow H$\alpha$ ($L_{H\alpha}$) 
in QSOs reported in \citet{gh05}, 
\begin{equation}
L_{H\alpha}=(5.25\pm0.02)\times10^{42}(\frac{L_{con}}{10^{44}{\rm erg~s^{-1}}})^{1.157\pm0.005}{\rm erg~s^{-1}}
\end{equation},
the total H$\alpha$ luminosity can be well calculated as $L_{H\alpha}\sim8.7\times10^{40}{\rm erg~s^{-1}}$. 
After considering the reddening corrected narrow H$\alpha$ luminosity of about $1.8\times10^{40}{\rm erg~s^{-1}}$, 
the broad H$\alpha$ luminosity could be $6.9\times10^{40}{\rm erg~s^{-1}}$, leading to the observed 
reddened broad H$\alpha$ flux to be about $666\times10^{-17}{\rm erg~s^{-1}~cm^{-2}}$. Spectroscopic 
properties with considerations of the expected broad H$\alpha$ can be re-constructed and shown in 
Fig.~\ref{fake} by the observed SDSS spectrum plus the broad Gaussian component with the central 
wavelength 6564\AA, the second moment from ${\rm 303km~s^{-1}}$ to ${\rm 996km~s^{-1}}$ and line 
flux $666\times10^{-17}{\rm erg~s^{-1}~cm^{-2}}$. It is clear that the expected Gaussian described 
broad H$\alpha$ was strong enough that the expected broad emission lines cannot be overwhelmed 
by noises in the observed spectrum of the \obj.

     The second method is applied as follows, similar as what have been done and discussed in 
\citet{av76, ly15}. Accepted the best fitted results $Y_{fit}$ to the emission lines around 
H$\beta$ and around H$\alpha$ by six narrow Gaussian functions shown in Fig.~\ref{line} with 
the determined $\chi^2/dof=200.446/231=0.87$, new values of $\chi^2_N/dof_N$ can be calculated 
after considering $Y_{fit}$ plus two another broad components for the broad H$\beta$ and the broad 
H$\alpha$. Here, not totally similar as what have been done in \citet{ly15}, not only the line flux 
$f_{line}$ but also the line width $\sigma_{line}$ of the two additional broad components are 
assumed to be free parameters. And we accept that the assumed broad H$\alpha$ and broad H$\beta$ 
have the same line width $\sigma_{line}$ which is larger than $400{\rm km~s^{-1}}$ and smaller 
than $3000{\rm km~s^{-1}}$ (the value is larger enough after considering the central BH mass in 
\obj), and have the flux ratio fixed to be 3.1. Then, after randomly collected 5000 data points 
of $\sigma_{line}$ and $f_{line}$, the dependence of $\chi^2_N/dof_N - 0.87$ on the input line 
flux $f_{line}$ of broad H$\beta$ shown in Fig.~\ref{chi2} which can be applied to determine the 
upper limits of line flux of broad H$\beta$ in the \obj. The 68\%, 90\% and 99\% confidence level 
upper limits of the line flux of broad H$\beta$ are determined as $46\times10^{-17}{\rm erg~s^{-1}~cm^{-2}}$, 
$82\times10^{-17}{\rm erg~s^{-1}~cm^{-2}}$ and $131\times10^{-17}{\rm erg~s^{-1}~cm^{-2}}$. 
Considering the reddening corrected continuum intensity about 
$9.7\times10^{-17}{\rm erg~s^{-1}~cm^{-2}~\textsc{\AA}^{-1}}$ underneath the H$\beta$ with the 
determined $E(B-V)=0.22$, the intrinsic equivalent width (EW) of broad H$\beta$ in the \obj~ should 
be smaller than 4.7\AA, 8.4\AA~ and 13.5\AA~ with 68\%, 90\% and 99\% confidence levels. Comparing 
with 90\% confidence level $EW<12.4$\AA~ of the broad H$\beta$ in the candidate of Type-2 AGN SDSS 
J0120 in \citet{ly15}, it is more reliable to confirm no broad optical emission lines in the \obj.

\subsection{\obj~ as a changing-look AGN at dim state?}

        In the subsection, we mainly consider whether the \obj~ should be a changing-look 
AGN at dim state?

	Changing-look AGN is one precious subclass of AGN with type transitions between 
Type-1 and Type-2. Since the first reported changing-look AGN NGC7603 in \citet{to76} with its 
broad H$\beta$ becoming much weaker in one year, there are dozens of changing-look AGN discovered, 
see the results well discussed in \citet{sb93, la15, mh16, mr16, gh17, rf18}, etc. And \citet{yw18} 
have reported a sample of 21 SDSS changing-look AGN with the appearance or the disappearance of 
broad Balmer emission lines within a few years. More recently, We \citet{zh21b} have reported a 
new changing-look quasar SDSS J2241. The main objective to consider properties of changing-look 
AGN is that changing-look AGN at dim state have disappearance of broad Balmer emission lines. 
Therefore, it is interesting to consider whether the \obj~ is actually a changing-look AGN at dim 
state.

	The dependence of AGN continuum luminosity on total \oiii~ line luminosity can be checked 
in \obj. For changing-look AGN with type transition within a few years, variability of central 
activities have strong effects on continuum emissions but few effects on narrow line emissions. 
Therefore, to check whether is there the same dependence of AGN continuum luminosity on total 
\oiii~ line luminosity in the \obj~ as in normal broad line AGN can provide clear clues to support 
or to rule out that the \obj~ is a changing-look AGN at dim state. The total \oiii~ line luminosity 
can be well measured as $3.2\times10^{39}{\rm erg~s^{-1}}$ in \obj. Considering the correlation 
between AGN continuum luminosity and total \oiii~ line luminosity shown in \citet{zh17}, 
\begin{equation}
\log(\frac{L_{con}}{{\rm erg~s^{-1}}})=(12.43\pm0.51)+
	(0.764\pm0.012)\times\log(\frac{L_{{\rm [O~\textsc{iii}]}}}{{\rm erg~s^{-1}}})
\end{equation},
the expected AGN continuum luminosity at 5100\AA~ should be around $4\times10^{42}{\rm erg~s^{-1}}$, 
well consistent with the estimated AGN continuum luminosity of ${\rm 2.89\times10^{42}erg~s^{-1}}$ 
in \obj. Properties of continuum luminosity and total \oiii line luminosity of \obj~ are shown 
in Fig.~\ref{co3}. Therefore, the \obj~ dose follow the same dependence of continuum luminosity 
on total \oiii~ line luminosity as normal broad line AGN, strongly indicating that the \obj~ is 
not a changing-look AGN at dim state, but is a better candidate of true Type-2 AGN.

\subsection{Which model is preferred to explain the nature of \obj?}

    As discussed in the Introduction, there are different theoretical models applied to explain 
the disappearance of central BLRs, such as the models with expected lower accretion rates and lower 
luminosities in \citet{eh09, cao10} and the models with expected higher luminosities in \citet{en16}.

    The bolometric luminosity of \obj~ can be estimated as
\begin{equation}
	L_{\rm bol}\sim10\times L_{\rm 5100}\sim3.9\times10^{43}{\rm erg~s^{-1}}
\end{equation}.
Here, we accept the optical bolometric correction $L_{\rm bol}\sim10\times L_{\rm 5100}$ which 
is mainly from the statistical properties of spectral energy distributions of broad line AGN 
discussed in \citet{rg06, du20} and from the more recent discussed results in \citet{nh20} 
based on theoretical calculations. It is clear that the bolometric luminosity of \obj~ is three 
magnitudes smaller than the critical value $4\times10^{46}{\rm erg~s^{-1}}$ in \citet{en16}, and 
also quite larger than the critical luminosity in \citet{eh09}.

\begin{figure}[htp] 
\centering\includegraphics[width = 9cm,height=6cm]{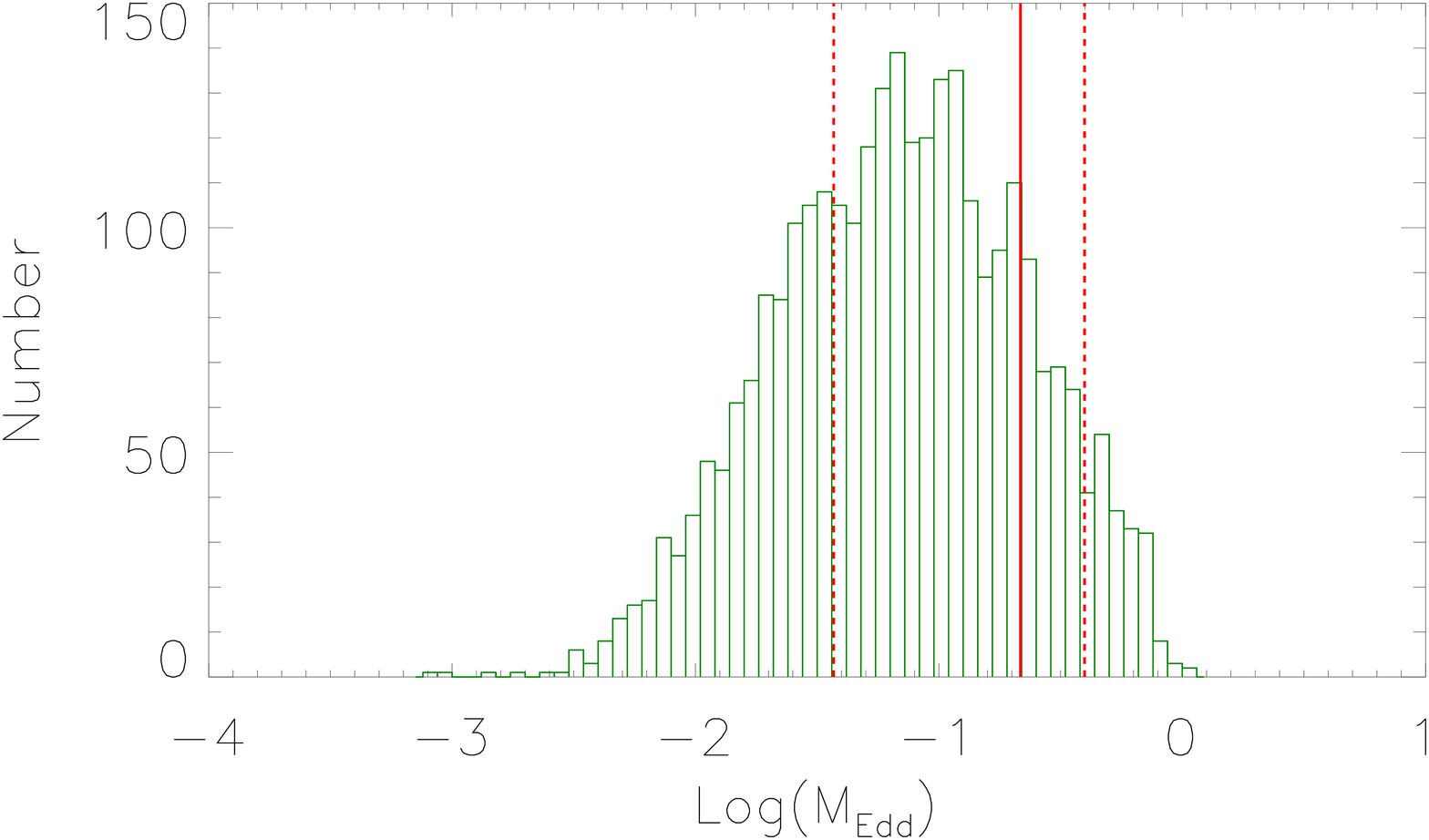}
\caption{Properties of Eddington ratio of \obj. Histogram shows the Eddington ratio distributions 
of the 2872 SDSS quasars from \citet{sh11}. Vertical solid red line marks the position 
$M_{\rm Edd}=0.216$, and vertical dashed red lines mark positions $M_{\rm Edd}=0.216\pm0.179$.}
\label{edd}
\end{figure}

     Meanwhile, based on the estimated minimum and maximum BH masses of \obj~ through the \msig 
relations and the estimated bolometric luminosity above, the dimensionless Eddington ratio 
$M_{\rm Edd}$ can be applied to trace intrinsic accretion rate in the \obj~ and estimated by
\begin{equation}
\begin{split}
M_{\rm Edd}&=\frac{L_{\rm bol}}{1.4\times10^{38}M_{\rm BH}/{\rm M_{\odot}}} \\
	& \sim0.216\pm0.179
\end{split}
\end{equation}.
Then, we compare the Eddington ratio of \obj~ and the normal SDSS quasars reported in \citet{sh11} 
in Fig.~\ref{edd}. The 2872 SDSS quasars ($z<0.33$) are collected from the sample in \citet{sh11} 
by three criteria, redshift less than 0.33, the measured continuum luminosity at least 10 times 
larger than the corresponding uncertainty and the measured BH mass at least 10 times larger than 
the corresponding uncertainty. The Eddington ratios of the 2872 SDSS quasars are calculated by the 
same equation above. It is clear that the estimated Eddington ratio of \obj~ is a common value among 
normal SDSS quasars.     	

	The central AGN activities of \obj~ are discussed above through the parameters of the 
continuum luminosity ${\rm 2.89\times10^{42}erg~s^{-1}}$ (or bolometric luminosity 
$3.9\times10^{43}{\rm erg~s^{-1}}$), and the Eddington ratio around 0.216. In the case of \obj~ 
as a candidate of true Type-2 AGN, not similar as the expected BLRs disappearance in most of 
candidates of true Type-2 AGN with expected quite lower continuum luminosity (and/or quite lower 
Eddington ratios) or with expected quite higher luminosities, the \obj~ has normal continuum 
luminosity relative to the [O~{\sc iii}] line luminosity and normal Eddington ratio, indicating 
\obj~ has unique properties among the reported true Type-2 AGN. And further efforts should be 
necessary to consider the loss of broad emission lines in the candidate of true Type-2 AGN \obj.

	The results discussed above can be well applied to give the conclusion that the \obj~ 
is a better candidate of true Type-2 AGN, however more further efforts in the near future are 
necessary to robustly confirm the conclusion. First and foremost, multi-epoch spectroscopic 
results are necessary to give further robust clues that \obj~ is not a changing-look AGN at dim 
state, besides the results shown in Fig.~7\ on properties of continuum luminosity and total 
[O~{\sc iii}] luminosity. Besides, multi-band observation results are necessary to give more 
accurate estimation of bolometric luminosity. Unfortunately, in the current stage, there are 
only photometric data points in optical band and near infrared band for \obj. Therefore, the 
bolometric correction is roughly applied in the manuscript. Last but not the least, independent 
method is necessary to be applied to measure central BH mass of \obj, in order to find more 
accurate Eddington ratio.

	It is clear that the \obj~ can be well confirmed as a better candidate of 
true Type-2 AGN but with unique properties among the reported true Type-2 AGN, especially 
its natural luminosity and natural Eddington ratio as normal quasars. The results probably 
indicate that some unique AGN with disappearance of central BLRs are not due to particular 
properties of central accretion processes but could be be treated as a subclass of Type-1 
AGN at special evolution stages. In the near future, some more true Type-2 AGN like \obj~
could provide further clues on the nature of true Type-2 AGN.

    Before the end of the section, simple discussions are listed on the pros and cons of 
the method to detect candidates of true Type-2 AGN by features of the lack of broad emission 
lines combining with the apparent long-term variability as what have been applied in the manuscript. 
The pros are as follows. First, the method applied in the manuscript is convenient to be applied 
to detect candidates of True Type-2 AGN among the large sample of narrow emission line galaxies 
in SDSS, combining with public light curves from CSS \citep{dr09}, ASAS-SN (All-Sky Automated 
Survey for Supernovae, \url{http://www.astronomy.ohio-state.edu/asassn/index.shtml}) \citep{sh14, 
ko17}, ZTF (Zwicky Transient Facility, \url{https://www.ztf.caltech.edu/}) \citep{bk19, ml19}, 
etc. Second, the method will lead to find more candidates of true Type-2 AGN with similar properties 
as those of normal broad line AGN, such as the \obj~ reported in the manuscript, which could 
provide further considerations on physical nature of true Type-2 AGN. Third, the method mainly 
focuses on spectroscopic emission line features and long-term variability properties, leading to 
conveniently detect candidates of true Type-2 AGN at high redshift. Certainly, there are cons as 
follows. First, the final results also depend on quality of spectrum. High quality spectrum should 
lead to different conclusion, such as the more recent results in NGC3147 with detected broad lines 
in high quality HST spectrum. Second, lack of multi-band properties of spectral energy distributions 
will lead to estimated bolometric luminosity and Eddington ratio not accurate enough. Third, 
the method sensitively depends on apparent long-term variability properties, leading to miss 
candidates of true Type-2 AGN with unapparent variability. 

\section{Conclusions}

   Finally, we give our main conclusions as follows. 
\begin{itemize}
\item Based on the high quality spectroscopic properties of \obj, a power law continuum component 
	is preferred with confidence level higher than 7sigma through the F-test technique, besides 
	the commonly considered host galaxy contributions.
\item Emission line fitting procedure provides robust evidence to support that there are no broad 
	Balmer emission lines in \obj. The F-test technique can tell the confidence level only around 
	6\% to support broad emission line components.
\item The upper limits of broad H$\beta$ line flux can be well estimated, leading the EW of 
	broad H$\beta$ to be smaller than 13.5\AA~ with 99\% confidence level, indicating no 
	intrinsic broad emission lines in the SDSS spectrum of \obj.
\item Under the Virialization assumption to BLRs combining with the strong correlation between 
	continuum luminosity and H$\alpha$ line luminosity, the expected broad H$\alpha$ (if 
	there was) can be well re-constructed, indicating that the expected broad H$\alpha$ 
	can not be overwhelmed by noises in the SDSS spectrum of \obj.
\item The long-term photometric variability of \obj~ can be well described by DRW process with 
	determined timescale about $100{\rm days}$, similar as the variability of normal SDSS 
	quasars, strongly supporting the central AGN activities.
\item \obj~ does follow the same dependence of continuum luminosity on total [O~{\sc iii}] line 
	luminosity as normal broad line AGN, indicating that \obj~ is not a changing-look AGN at 
	dim state but a better candidate of true Type-2 AGN.
\item \obj~ has normal luminosity and Eddington ratio, not similar as the expected lower 
	luminosities (lower accretion rates) or higher luminosities, indicating \obj~ is an unique 
	Type-2 AGN with lack of central BLRs.
\end{itemize}

\section*{Acknowledgements}
We gratefully acknowledge the anonymous referee for carefully reading our manuscript with patience 
and giving us constructive comments and suggestions to greatly improve our paper. Zhang gratefully 
acknowledges the kind support of Starting Research Fund of Nanjing Normal University and from the 
financial support of NSFC-11973029. This manuscript has made use of the data from the SDSS projects. 
The SDSS-III web site is http://www.sdss3.org/. SDSS-III is managed by the Astrophysical Research 
Consortium for the Participating Institutions of the SDSS-III Collaboration. 

\end{document}